\documentclass[10pt,twocolumn]{article}
\usepackage{isqcmc2025} 
\usepackage{changepage}
\usepackage{graphicx,url}
\usepackage{amsmath}
\usepackage{url}
\usepackage{hyperref}

\usepackage{ulem}
\usepackage[font=bf, labelfont=bf, justification=justified, format=plain, margin=0.8cm]{caption}

\usepackage[utf8]{inputenc}

\usepackage{lmodern}  

\usepackage{anonymize} 
\usepackage{listings}

\lstdefinelanguage{OPENQASM}{
morekeywords={q, c}, 
morekeywords=[2]{OPENQASM, include}, 
emph={h,cx,qreg,creg,->}, 
morekeywords=[3]{measure}, 
morekeywords=[4]{barrier}, 
sensitive=true,
morecomment=[l]{//}, 
morestring=[b]",
literate={->}{{\textbf{\color{codeemph2}{$\to$}}}}1 
}

\title{Sonification of entanglement dynamics in many-qubit systems}

\author{\anonymize{Juliette Tudoce}\inst{1} \and \anonymize{Marcin Płodzień}\inst{1} \and
        \anonymize{Maciej Lewenstein}\inst{1}\inst{2} \and
        \anonymize{Reiko Yamada}\inst{1}\inst{3} 
        }

\address{\anonymize{ICFO-Institut de Ciencies Fotoniques, The Barcelona Institute of Science and Technology}\\
         \anonymize{Av. Carl Friedrich Gauss 3, 08860 Castelldefels (Barcelona), Spain}
         \nextinstitute
         \anonymize{ICREA, Pg. Lluís Companys 23, 08010 Barcelona, Spain}
         \nextinstitute
         \anonymize{ESMUC, Edifici de L'Auditori, C/ de Padilla, 155, Eixample, 08013 Barcelona, Spain}
}

\begin{document}

\maketitle

\begin{abstract}

Quantum mechanics poses significant challenges for audio-visual representation, particularly concerning quantum entanglement. Sonification---the auditory representation of data---offers a promising complementary approach. This paper investigates sonification techniques applied to dynamical entanglement generation in many-qubit systems with the help of phase space methods and entanglement measure. 
We study dynamics of entanglement generation in many-qubit system in dynamical protocol governed by two models: the one-axis twisting model, and a quantum kicked-rotor exhibiting both regular and quantum chaotic behavior. We present a procedure of entanglement dynamics sonification, allowing mapping the phase-space representation of a  many-qubit quantum state and von Neuman entanglement entropy to sound.
Results demonstrate how sonification enhances perception of dynamic entanglement offering intuitive and artistic insight into quantum correlations behaviors.
\end{abstract}

\section{Introduction}\label{sec:introduction}

Fundamental, microscopic description of reality requires using formalism of quantum mechanics, where quantum states describing properties of a given system live in a highly-dimensional vector Hilbert space.
Quantum theory predicts phenomena which do not have classical counterparts. These phenomena are manifested in quantum correlations, such as entanglement and Bell correlations, which are fundamental features of many-qubit systems~\cite{horodecki2009entanglement,brunner2014bell,Srivastava2024,horodecki2024multipartiteentanglement}.

The {\it First Quantum Revolution} in the XX century was related to reformulating our understanding of reality in the light of consequences of fundamental aspects of quantum mechanics prediction. Three decades ago, we entered the {\it Second Quantum Revolution}, since experimental techniques allow for precise preparation and control of quantum systems. The high level of control over the quantum dynamics allows to generate, and store quantum states possessing quantum correlations, utilized later for information processing. These quantum correlations are fundamental resources for future quantum-based technologies, namely quantum computing, quantum communication, quantum simulation, and quantum sensing \cite{Altman2021, Osada2022, Fraxanet2023, Simon2025}.

Over the recent years, next to fundamental research on quantum theory, as well to research on quantum-based technologies, quantum mechanics attracts the interest of the artist community. From an artistic point of view, one of the main challenges is to provide an audio-visual representation of quantum phenomena. Quantum mechanical concepts are well defined in the proper mathematical framework, however, they are hard to understand at the level of everyday intuition - as they do not have any classical counterpart. Nevertheless, in the field of computer music, and by extension electroacoustics, there is a constant effort to build the route allowing to travel from formal mathematical language of quantum mechanics, to audio-visual interpretation of non-classical reality \cite{Miranda:2025}. 

The human auditory system excels at recognizing patterns, temporal structures, and subtle variations over time, making sonification a powerful tool for interpreting intricate datasets. 
Sonification provides a
route for rendering otherwise abstract physical processes—including quantum phenomena—into audible form. As Kramer et al. (1999) put it, “Sonification is the transformation of data relationships into perceived relationships in an acoustic signal for the purpose of facilitating communication or interpretation” \cite{Kramer:1999}. Although the technique is not new—the clicking Geiger–Müller counter dates to 1908—its adoption in complex scientific domains has grown markedly over the past few decades \cite{Flowers1995,Cooke2017,Scaletti2018}. Modern applications span mechanical engineering~\cite{Krishnan2001-bu,Hildebrandt2014-ds,Iber2021-mc}, oceanography~\cite{Sturm2005,Smith2023}, seismology~\cite{Tang2014,Pat2016,Boschi2017,Pat2021}, and astronomy~\cite{Misdariis2022,Harrison2024}. Perhaps the most spectacular example to date is the direct detection—and subsequent audible “chirp”—of gravitational waves from a binary‑black‑hole merger \cite{Abbott2016-ip,Rebollo-Neira2018-av}.

In the field of quantum physics, sonification remains relatively underexplored in scientific contributions.
Over the past thirty years, sonification has been applied to a variety of single particle quantum problems, including particle motion based on de Broglie's hypothesis \cite{Sturm:2000}, Gaussian wave packets in potential wells \cite{Cadiz_Ramos:2014}, quantum oscillators \cite{DuPlessis:2021, Yamada:2023}, and molecular electronic energy densities in quantum chemistry \cite{Arasaki_Takatsuka:2024}. The new challenge for sonification of quantum mechanics lies in perceptual representation of many-body quantum systems dynamics, where Hilbert space size grows exponentially with number of particles, and quantum correlations emerge, contrary to single particle quantum mechanical systems, where only quantum coherence can appear.

In this work we tackle the problem of sonification of dynamics in many-body quantum system.
We focus on dynamical quantum entanglement generation in spin-$1/2$ chains, starting from an initial product state. We consider two models: one-axis twising (OAT), and quantum kicked rotor. 
With the help of phase space methods, and quantum information theory, we present sonification procedure allowing to 
construct visual-sonic bridge, translating abstract dynamics into perceptible features.

The paper is organized as follow:
Section~\ref{sec:background} provides the background on generation of quantum entanglement in dynamical protocols. Section~\ref{sec:methodology} describes the methodology employed to map the data to sound. Section~\ref{sec:results} presents and analyzes the simulation results. Section~\ref{sec:discussion} discusses implications and outlines future directions, and Section~\ref{sec:conclusion} concludes the paper.

\section{Preliminaries}\label{sec:background}

This section presents the key tools relevant to our study. We define bipartite entanglement entropy, the Husimi $Q$ function, and present considered Hamiltonians governing the system dynamics and their traditional visualization. The explanations aim to balance clarity and rigor, providing essential context for both general readers and those with a background in quantum physics.

\subsection{Dynamical generation of quantum correlations}\label{subsec:dynamical_protocol}

The general protocol for dynamical generation of many-body quantum correlations starts with preparation of an initial state of the system in a product state $|\psi_0\rangle$. Next, the initial state evolves in time under interacting Hamiltonian $\hat{H}$, $
    |\psi(t)\rangle =  U_t|\psi_0\rangle
$, where $U_t = e^{-i t{H}}$ is time evolution operator. The interactions between spins transform initial uncorrelated state to entangled state.

\subsection{Entanglement entropy}\label{subsec:entanglement}

Quantum entanglement can be quantified in several inequivalent ways, with the appropriate choice of measure depending on factors such as whether the global state is pure or mixed, the number of parties involved, and the operational task under consideration \cite{Horodecki:2009,Bengtsson2006,Plenio2014,Ma2024}. Because the states we study are globally pure and we are interested in a single bipartition of the system, we adopt the bipartite von Neumann entropy $S_{vN}$.

The bipartite entanglement entropy can be obtained from properties of a reduced density matrix of a system after partial measurement. Considering an arbitrary division of a spin-$1/2$ chain of length $L$ into two subsystems $A$, and $B$, the reduced density matrix of the subsystem $A$ is defined as 
 $\rho_A(t) = \text{Tr}_B[\rho(t)]$,
where $\rho(t) =|\psi(t)\rangle\langle \psi(t)| = U_t|\psi_0\rangle\langle\psi_0|U^\dagger_t$ is a density matrix of time evolved state, while trace operation represents measurements of spins belonging to $B$-subsystem (without loss of generality we will consider even $L$, and equal division of the system, $L_A = L_B = L/2$).
The von Neuman Entropy $S_{vN}$ definition reads:
\begin{equation}\label{eq:SvN}
    S_{vN}(t)
    = -\sum_i\lambda_i(t) \log(\lambda_i(t)),
\end{equation}
where $\lambda_i(t)$ are eigenvalues of reduced density matrix $\rho_A(t) = \sum_i\lambda_i(t)|\lambda_i(t)\rangle\langle\lambda_i(t)|$, at given time $t$. 
Bipartite entropy is a bounded quantity $0\le S_{vN}(t)\le \log_2({\rm dim}{\cal H}_A)$, where ${\rm dim}{\cal H}_A$ is a Hilbert space size of the subsystem A. It vanishes for 
separable states, and takes maximal value for a quantum state being a product of $L/2$ Bell pairs, i.e. $|\psi\rangle = |\Phi_+\rangle^{\otimes L/2}$, where $|\Phi_+\rangle=\frac{1}{\sqrt{2}}(|00\rangle + |11\rangle)$.

\subsection{Husimi Q Function}\label{subsec:husimiq}

An exact visualization of a many‑body quantum state is out of reach, because the dimension of its Hilbert space scales exponentially with number of spins. 
A convenient approximate picture, however, can be obtained by projecting the state onto spin‑coherent states. A spin‑coherent state $|\theta, \phi\rangle$ represents an ensemble of product of $L$ spins collectively pointing in the same direction specified by the spherical polar angles $(\theta,\phi)$, and is defined as
\begin{equation}
  \lvert\theta,\phi\rangle
  = e^{-i\phi S_z}\,e^{-i\theta S_y}\,
    \lvert0\rangle^{\otimes L},
  \label{eq:SCS}
\end{equation}
where $|0\rangle^{\otimes L}$ is the product state in which every spin is in spin-up configuration, $|0\rangle\equiv|\uparrow\rangle$.
The collective spin operators $S_{y,z} = \frac{1}{2}\sum_{i=1}^{L}\sigma^{y,z}$, where $\sigma^{y,z}_i$ are Pauli operators acting on $i$-th spin,
generate global rotations of this product state:
\(e^{-i\theta S_y}\) rotates the spins by the polar angle \(\theta\) about the \(y\)-axis, and \(e^{-i\phi S_z}\) subsequently rotates them by the azimuthal angle \(\phi\) about the \(z\)-axis.

An approximate visualization of spin-$1/2$ chain quantum state $|\psi\rangle$ can be obtained via spin-$1/2$ chain version of a Husimi function $Q(\theta,\phi)$, defined as
$Q(\theta, \phi) = |\langle \theta, \phi|\psi\rangle|^2$.
The Husimi $Q$ function represents probability density
that $|\psi\rangle$ is in a spin-coherent state $|\theta, \phi\rangle$, where all spins in state $|\psi\rangle$ collectively point the same direction defined by angles $(\theta, \phi)$.

For dynamical protocols, a Husimi function has to be specified at each time $t$, i.e.
\begin{equation}\label{eq:Husimi_Q_t}
    Q_t(\theta, \phi) = |\langle \theta,\phi|\psi(t)\rangle|^2 = |\langle \theta,\phi|e^{-i t H}|\psi_0\rangle|^2,
\end{equation}
and its change in time provides an insight into the dynamics of the many-qubit system. When Husimi function takes values close to $1$ for a given coordinates $(\theta_*,\phi_*)$ then interpretation of the quantum state $|\psi\rangle$ is clear - system has small entanglement (or zero if $Q_t(\theta_*,\phi_*) = 1$, and all spins point in the same direction, defined by spherical angles. However, there is no clear interpretation of the structure of the quantum state $|\psi\rangle$, when a Husimi is close to $0$ for a given pair $(\theta_*,\phi_*)$: we can only say that it has small (or vanishing) overlap with a coherent state $|\theta_*,\phi_*\rangle$.

Because Husimi function alone does not provide insight into the amount of entanglement in the system, in the following sections, we use both the entanglement entropy $S_{vN}(t)$ and Husimi function $Q_t(\theta, \phi)$ to provide audio-visual characterisation of the many-qubit system dynamics capturing quantum correlations.

\subsection{Models}

In this work we consider two interacting Hamiltonians which dynamically generate quantum correlations, starting from the initial spin coherent state $|\psi_0\rangle = e^{-i\phi_0 S_z}e^{-i\theta_0 S_y}|0\rangle^{\otimes L}\equiv|\theta_0, \phi_0\rangle$.

\subsubsection{One-axis twisting}

\begin{figure}[t!]
    \centering
    \includegraphics[width=.45\textwidth]{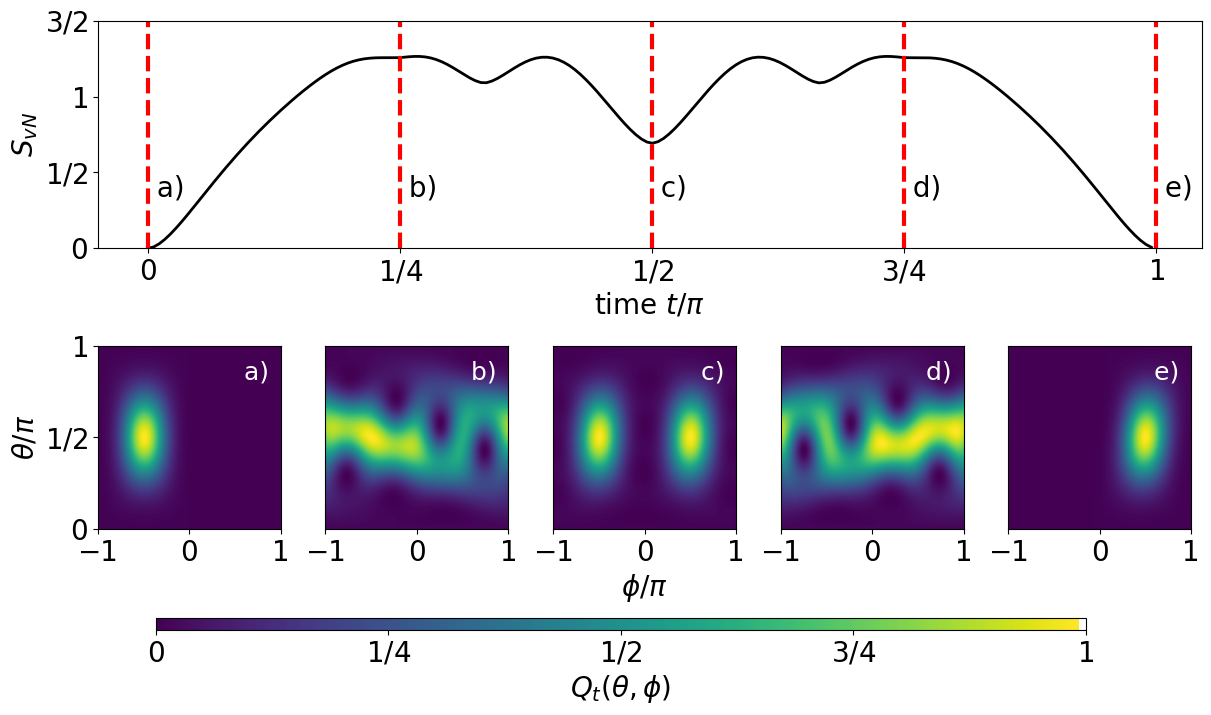}
    \caption{Top panel presents time evolution of the entanglement entropy for $L=8$ spins under the OAT evolution with an initial state product state $|\psi_0\rangle = |\pi/2,-\pi/2\rangle$.
Bottom panels (a-e) present Husimi $Q_t(\theta, \phi)$ function at different times $t = 0, \pi/4, \pi/2, \pi$ (panels (a-e) respectively). Vertical red-dashed lines indicates value of the entanglement entropy at given times, corresponding to presented Husimi functions. At time $t = \pi/2$ the $L$-body GHZ state is generated.}
    \label{fig:fig1}
\end{figure}

The first considered model is the One-Axis Twisting Hamiltonian, defined as $H  = S_z^2 = \frac{1}{4}\sum_{i<j}\sigma^z_i\sigma^z_j$, being a non-linear Hamiltonian with all-to-all couplings.
The OAT protocol dynamically generates metrologically useful spin-squeezed states ~\cite{kitagawa1993squeezed,wineland1994squeezed,RevModPhys.90.035005}, many-body entangled and the many-body Bell correlated states~\cite{Tura1256, Plodzien2022}. The OAT dynamics can be realized with modern quantum simulator platforms, such as ultra-cold atoms in optical lattices with bosonic or fermionic spieces \cite{Plodzien2020, HernndezYanes2022, Dziurawiec2023, HernndezYanes2023, Plodzien2024}.

Fig~\ref{fig:fig1} presents time-evolution of the Husimi function for $L=8$ spins initially prepared in a spin-coherent state $|\psi_0\rangle = |\pi/2,-\pi/2\rangle$, where all spins points in the same directions, and do not share entanglement ($S_{vN} = 0)$. During the evolution, quantum entanglement in the system is generated ($S_{vN}>0$), and at time $t = \pi/2$ a many-body GHZ (Greenberger–Horne–Zeilinger) state is generated, $|\psi\rangle = \frac{1}{\sqrt{2}}(|\pi/2, -\pi/2\rangle + |\pi/2,\pi/2\rangle)$ being a coherent superposition of all spins pointing in oposite directions along $x$-axis.

\subsubsection{Quantum Kicked Rotor}
We consider the quantum kicked rotor model \cite{Haake_Kus_Scharf:1987}, whose evolution is defined via the iterative map
$|\psi_{k+1}\rangle = U_\alpha V_\beta |\psi_k\rangle,
U_\alpha = e^{-i\frac{\alpha}{L} S_z^2},
V_{\beta} = e^{-i\beta S_y}$.
The dynamics is governed by the free rotation about $y$-axis generated by collective $S_y$ spin operator, while in a regular time-interval systems experiences the kick $S^2_z$ of strength $\alpha$. The quantum kicked rotor model exhibits regular-to-chaos transition \cite{Haake_Kus_Scharf:1987} depending an initial state $|\psi_0\rangle$, free evolution parameter $\beta$, and a kick strength $\alpha$. For small $\alpha <1$ system exhibits regular dynamics, while $\alpha > 1$ generates quantum chaotic behavior.
The parameters $\alpha/L$, and $\beta$ can be equivalently interpreted as duration of dynamics governed by $S^2_z$, or $S_y$ operator, respectively. In particular for $\beta = 0$, and initial state $|\psi_0\rangle = |\pi/2,\phi\rangle$, $\phi\in (-\pi, \pi)$, the quantum kicked rotor dynamics is equivalent to the OAT model.
\begin{figure}[t!]
    \centering
    \includegraphics[width=.45\textwidth]{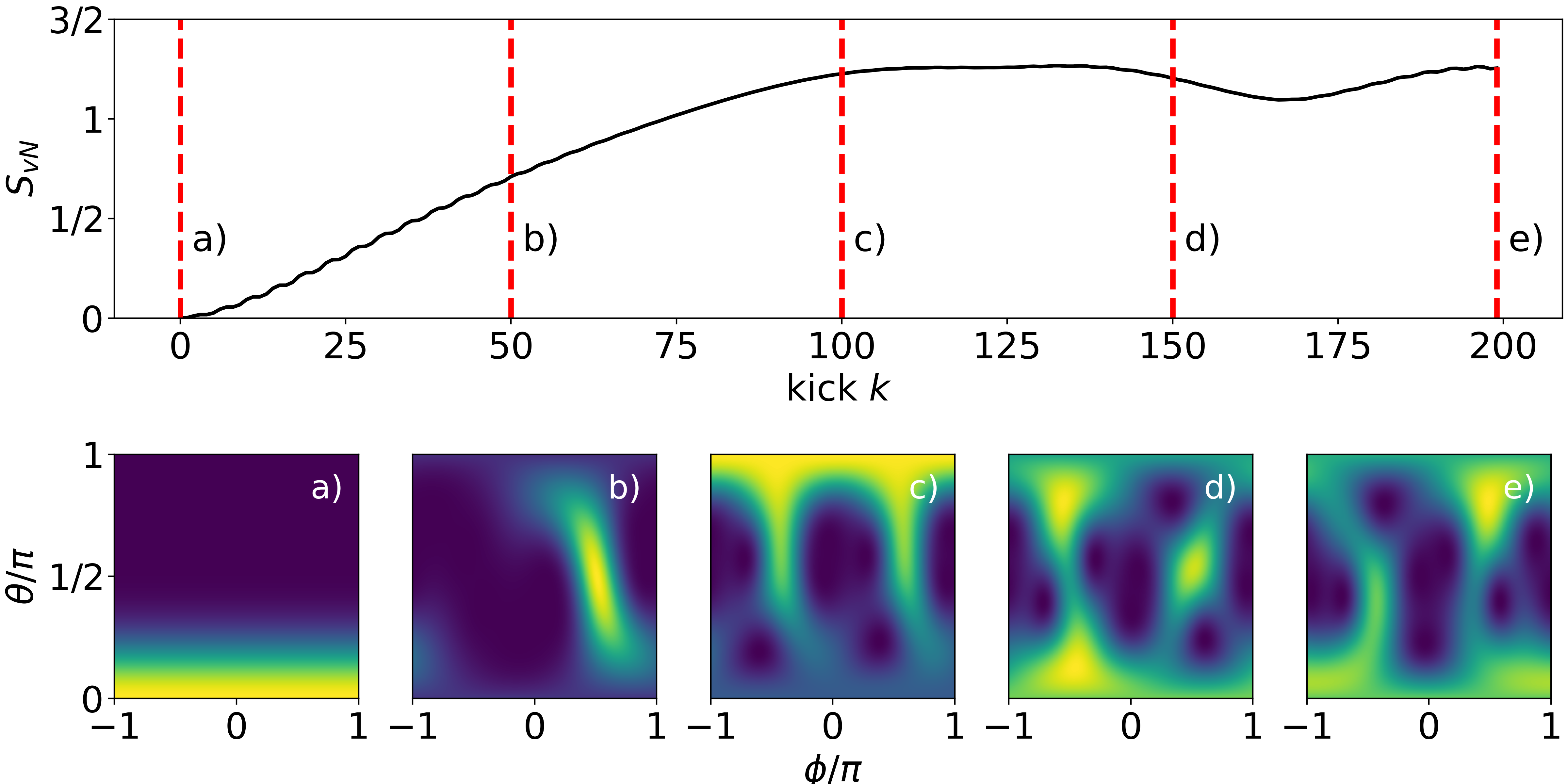}
    \caption{Top panel: entanglement entropy $S_{vN}$ evolution after $k$ kicks for the quantum kicked rotor with $\beta = \pi/2$ and $\alpha = 0.1$, with an initial state $|\psi_0\rangle = |0,0\rangle$. Bottom panel presents corresponding Husimi function $Q_t(\theta,\phi)$ after $k=0,50,100,150,200$ kicks (panels (a-e) respectively). Horizontal red dashed lines present $S_{vN}$ for chosen $k$. System undergoes a regular dynamics, which can be monitored via smooth evolution of $S_{vN}$.}
    \label{fig:fig2}
\end{figure}

\begin{figure}[t!]
    \centering
    \includegraphics[width=.45\textwidth]{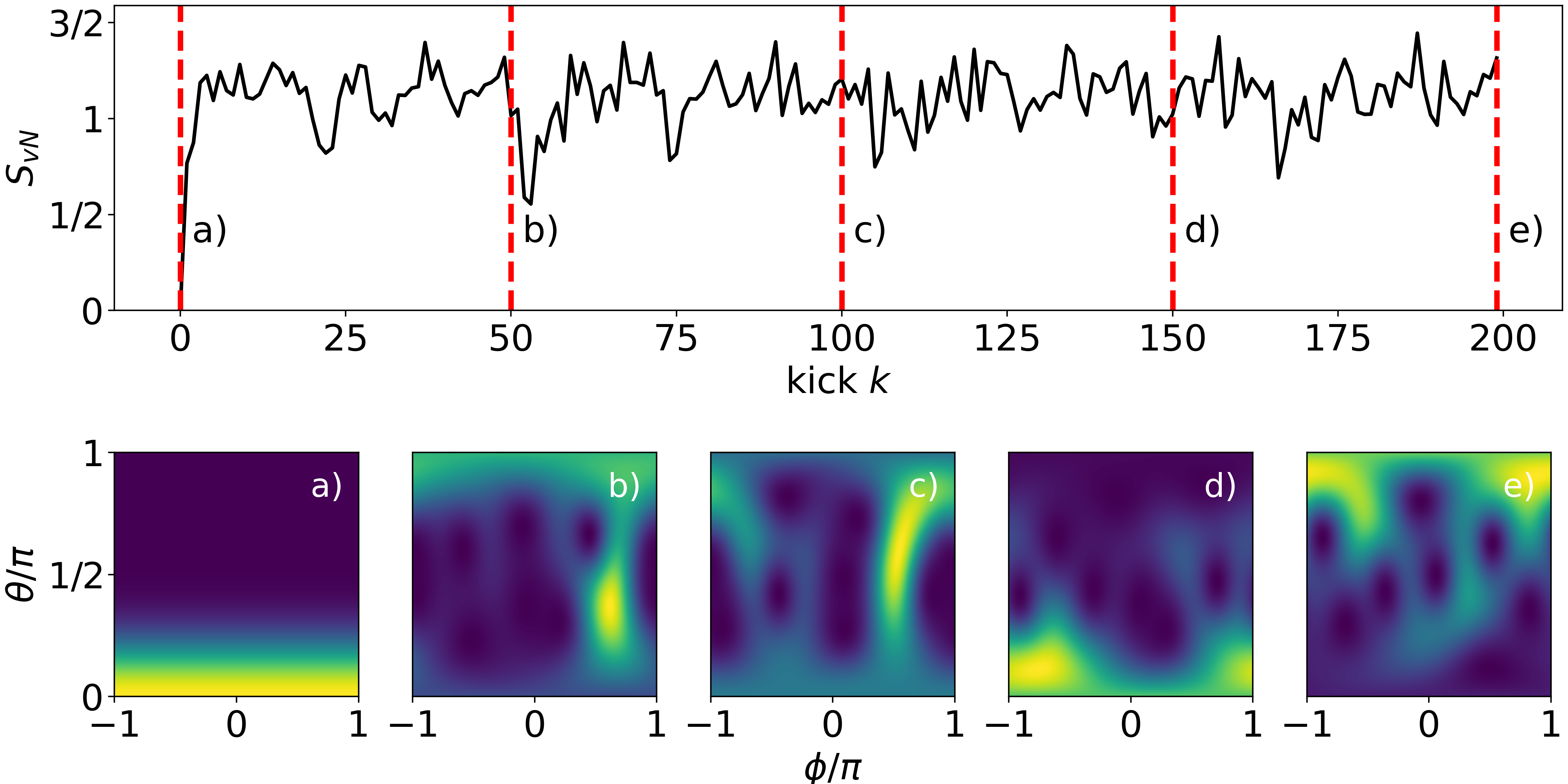}
    \caption{Same as Fig.~\ref{fig:fig2}, with $\alpha = 10$. The system is in a crossover regime between regular and chaotic dynamics, which can be monitored via rapid fluctuations of $S_{\mathrm{vN}}$.}
    \label{fig:fig3}
\end{figure}

In the following we choose an initial state to be $|\psi_0\rangle = |0,0\rangle$, i.e. all spins point to the north pole on the Bloch sphere, and we set $\beta = \pi/2$. We consider two values of kick strengths parameter, $\alpha = 0.1, 10$, corresponding to regular-, and chaotic dynamics, respectively. Evolution of the system is presented on Figs.\ref{fig:fig2},\ref{fig:fig3}. For a regular dynamics the entanglement entropy smoothly evolves with kicks, as well as corresponding Husimi function ($\alpha = 0.1$, Fig.\ref{fig:fig2}). System starts with product state, and during the evolution entanglement is smoothly generated. When the system is in a quantum chaotic regime with strong kick strength ($\alpha = 10$, Fig.\ref{fig:fig3}), entanglement entropy grows rapidly after few first kicks, and fluctuates around constant value.

\section{Methodology}\label{sec:methodology}

We propose a sonification procedure which transforms the Husimi function $Q_t(\theta,\phi)$ and the bipartite entanglement entropy $S_{vN}(t)$ during quantum evolution into auditory features. This mapping tries to represent these parameters in the most intuitive manner. The following subsections detail our sonification framework.

\subsection{Amplitude Mapping}\label{subsec:amplitude}
The amplitude (loudness) of the audio signal directly maps the amplitude of the probability density given by the Husimi Q function $Q(\theta,\phi)$. This choice reflects the fundamental quantum mechanical interpretation of $Q(\theta,\phi)$ as a quasi-probability distribution in phase space \cite{Husimi:1940}. The mapping follows
$A(t) \propto Q(\theta(t), \phi(t))$
where $A(t)$ represents the time-dependent amplitude. This proportional relationship ensures that regions of high quantum probability density produce more audible signals, while low-probability regions become perceptually negligible. The logarithmic nature of human loudness perception \cite{Moore:2012} makes this mapping particularly effective for distinguishing significant quantum features.

\subsection{Stereo Spatialization}\label{subsec:spatialization}

We employ the azimuthal angle $\phi$ to control stereo panning, creating a spatial representation of quantum phase space, where panning is proportional to 
$\phi$, i.e., $\text{panning} \propto \phi$, translating the angular coordinate into a left-right audio distribution. This approach mirrors the spatial symmetry of the quantum system, with $\phi = 0$ corresponding to the left channel and $\phi = \pi$ to the right channel. The linear proportionality maintains an intuitive correspondence between quantum phase space and auditory space. The sonification could be further enhanced by rendering it in a domed ambisonic environment, which would enable a more faithful spatial mapping of the full azimuthal ($\phi$) range.

\subsection{Pitch Encoding}\label{subsec:pitch}

Frequency (pitch) is determined by the polar angle $\theta$ relative to a reference initial state. Using $f_{\text{init}} = 440$ Hz as our reference frequency (A4), corresponding to the angle $\theta_{\text{init}}$ at which the system is prepared, we implement:
\begin{equation}
f = f_{\text{init}} - |\theta - \theta_{\text{init}}| \times f_{\text{init}} \times \text{spread}
\end{equation}
This mapping generates pitch deviations that correspond to angular displacements on the Bloch sphere. The modulation factor reflects the distributional spread of the Husimi Q function across the sphere. When regions of high probability density are more broadly distributed, the resulting frequency shifts are more pronounced. This approach makes pitch sensitive not only to directionality, but also to how widely the state is supported in phase space, offering a perceptual cue for quantum state delocalization \cite{Sakurai_Napolitano:2017}.

\subsection{Timbre Representation of Entanglement}\label{subsec:timbre}

Waveform complexity serves as an auditory representation of entanglement entropy $S_{vN}(\rho)$. We implement a continuous timbre space where:

\begin{itemize}
\item Minimal entropy ($S_{vN} \approx 0$) produces pure sinusoidal tones, corresponding to separable, unentangled quantum states.
\item Maximal entropy generates complex, harmonic-rich waveforms incorporating features such as triangle waves and ring modulation. 
\item Intermediate entropy levels are represented by an interpolation between pure sinusoidal and complex waveforms, with the proportion of harmonic complexity increasing continuously as the entropy rises.
\end{itemize}

Ring modulation involves multiplying two sine waves—an input tone and a carrier oscillator—resulting in a signal composed solely of the sum and the difference of the two frequencies. Mathematically, this is expressed as:
\begin{align}
\sin(2\pi f_1 t) \cdot \sin(2\pi f_2 t) &= \frac{1}{2} \big[ \cos\big(2\pi(f_1 - f_2)t\big) \nonumber \\
&\quad - \cos\big(2\pi(f_1 + f_2)t\big) \big]
\end{align}
where \( f_1 \) and \( f_2 \) are the input and carrier frequencies, respectively. This process eliminates the original frequencies and produces a metallic, bell-like timbre that is highly sensitive to frequency variation and phase relationships.

While the von Neumann entropy cannot be directly represented on the Bloch sphere—requiring separate plots to visualize its temporal evolution—our sonification approach provides a way to capture its dynamics in real time through auditory perception. By mapping the increasing entropy to progressive changes in waveform complexity, the method intuitively mirrors the growth of quantum correlations and the rising structural complexity of the quantum state \cite{Nielsen_Chuang:2010}. This auditory encoding enables simultaneous exploration of both the system’s phase-space structure and its entanglement dynamics, offering a multidimensional understanding of quantum behavior.

\section{Results}\label{sec:results}
To investigate the relationship between entanglement dynamics and auditory perception, we analyzed the spectrograms and von Neumann entropy (SvN) evolution for different Hamiltonians and system sizes. In our analysis, we focus on two complementary waveshapes for sonification: triangular waveforms and ring modulation. The triangular waveform introduces a rich harmonic structure, often perceived as higher-pitched and spectrally rich. Among standard periodic waveforms such as square, sawtooth, and absolute sine, the triangle wave was chosen for its perceptual balance of harmonic content—providing sufficient upper harmonics for spectral richness without overwhelming the signal with high-frequency energy. This selection was based on a subjective evaluation of clarity and interpretability in the resulting sound.
In contrast, ring modulation—despite producing fewer harmonics—yields a more metallic and textured timbre. Its strong sensitivity to amplitude variation results in perceptible silences when the von Neumann entropy is at its maximum and louder output when the entropy is low. This creates a pronounced dynamic range, enhancing the auditory contrast between weakly and strongly entangled states. Taken together, the two waveshapes highlight different aspects of the quantum dynamics, providing complementary perceptual perspectives on structure, complexity, and temporal evolution.

\subsection{One-Axis Twisting}\label{sec:results_OAT}

\begin{figure*}[htbp]   
  \centering
  \includegraphics[width=\textwidth]{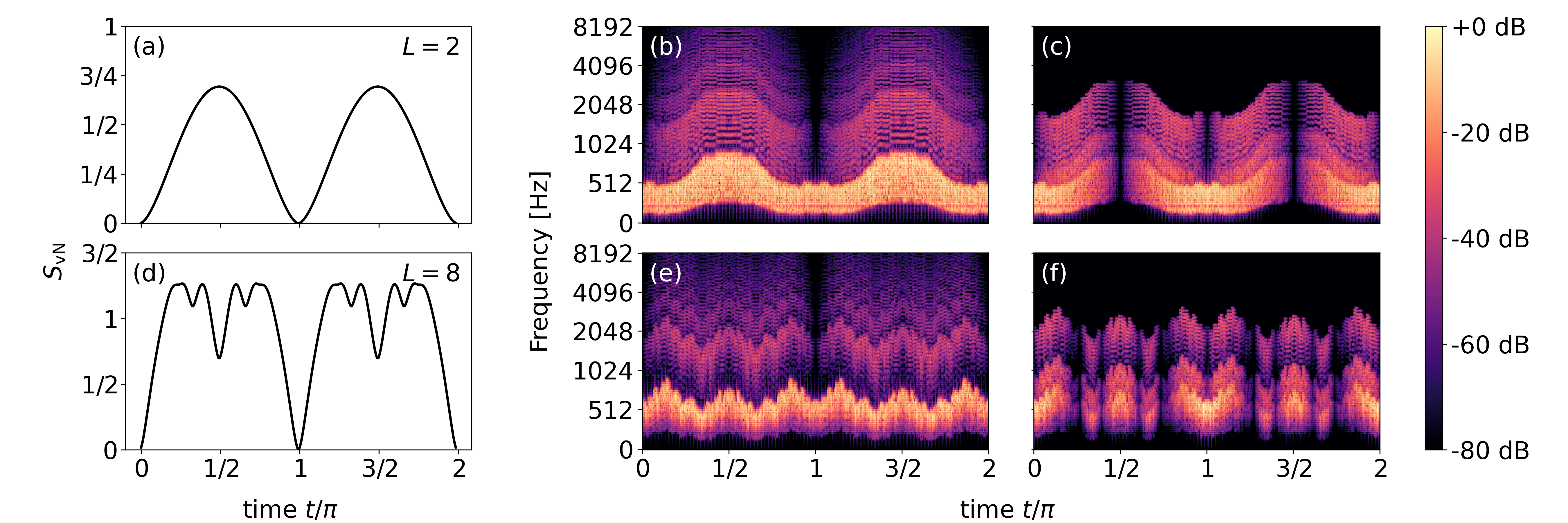}
  \caption{For the OAT Hamiltonian at system sizes \(L = 2\) [panels (a–c)] and \(L = 8\) [panels (d–f)].
           (a,d) von Neumann entropy \(S_{\text{vN}}\) evolution;
           (b,e) spectrograms obtained with a triangular waveform;
           (c,f) spectrograms obtained with ring modulation.
  Animations with sound are available for
  \href{https://youtube.com/shorts/vcfbH3STS-c}{panel~(b)},
  \href{https://youtube.com/shorts/sl5TPmSGk5Q}{panel~(c)},
  \href{https://youtube.com/shorts/iLcgV74upcY}{panel~(e)},
  and \href{https://youtube.com/shorts/bA5y5Q0b_hs}{panel~(f)}.}
  \label{fig:svn_spectrogram_OAT}
\end{figure*}

The OAT Hamiltonian produces smooth and periodic spectro-temporal patterns (see Fig.~\ref{fig:svn_spectrogram_OAT}). For \( L = 2 \), the spectrograms in panels (b) and (c) show broad, slowly evolving harmonic structures, which correspond to the sinusoidal oscillations in the von Neumann entropy shown in panel (a). For \( L = 8 \), panels (e) and (f) depict a richer yet still regular spectro-temporal texture, reflecting the more structured but smooth entropy dynamics in panel (d). Sonically, this results in gradually evolving textures characterized by a clear sense of periodicity. 

\subsection{Quantum Kicked Rotor}\label{sec:results_kicked}

\begin{figure*}[htbp] 
  \centering
  \includegraphics[width=\textwidth]{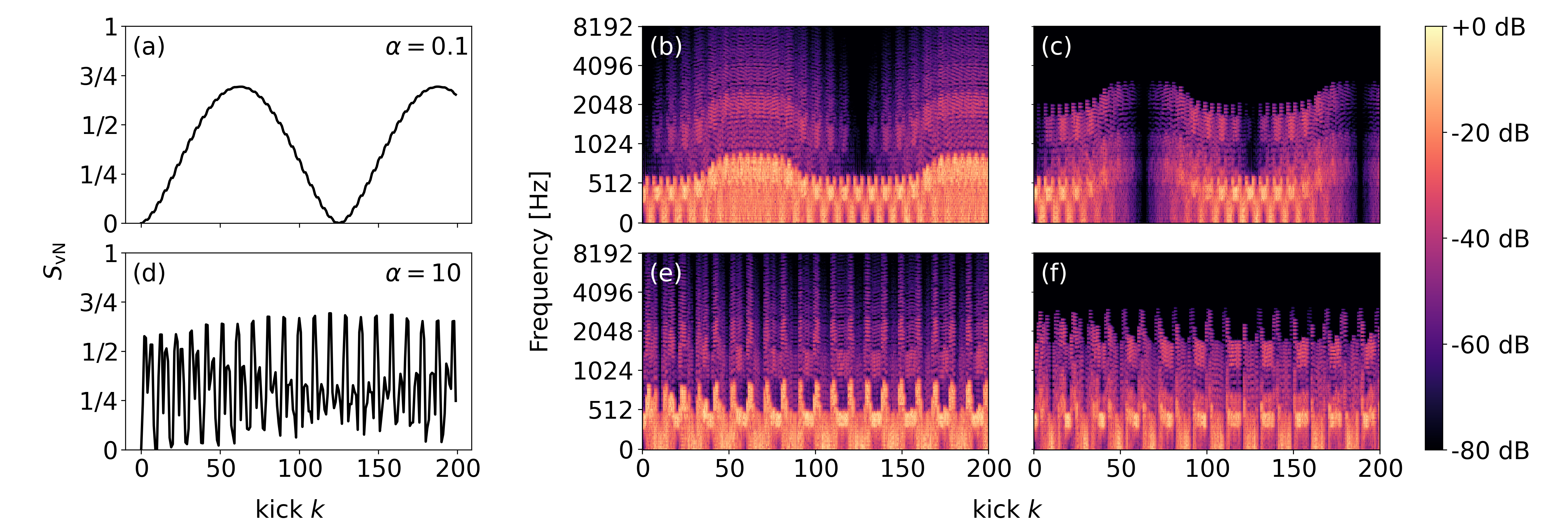}
  \caption{For the kicked rotor at system size \(L = 2\), prepared in the initial state \(\lvert\theta,\phi\rangle = \lvert 0,0\rangle\).
  Results are shown for \(\alpha = 0.1\) [panels (a–c)] and \(\alpha = 10\) [panels (d–f)].
  Panel labels correspond to those in Fig.~\ref{fig:svn_spectrogram_OAT}
    Animations with sound are available for
  \href{https://youtube.com/shorts/AeeVuEGamZc}{panel~(b)},
  \href{https://youtube.com/shorts/eG_-_DljTmA}{panel~(c)},
  \href{https://youtube.com/shorts/HDPLa6cCksY}{panel~(e)},
  and \href{https://youtube.com/shorts/LDwdGLdQfoA}{panel~(f)}.}
  \label{fig:svn_spectrogram_kicked_th0_L2}
\end{figure*}

\begin{figure*}[htbp]   
  \centering
  \includegraphics[width=\textwidth]{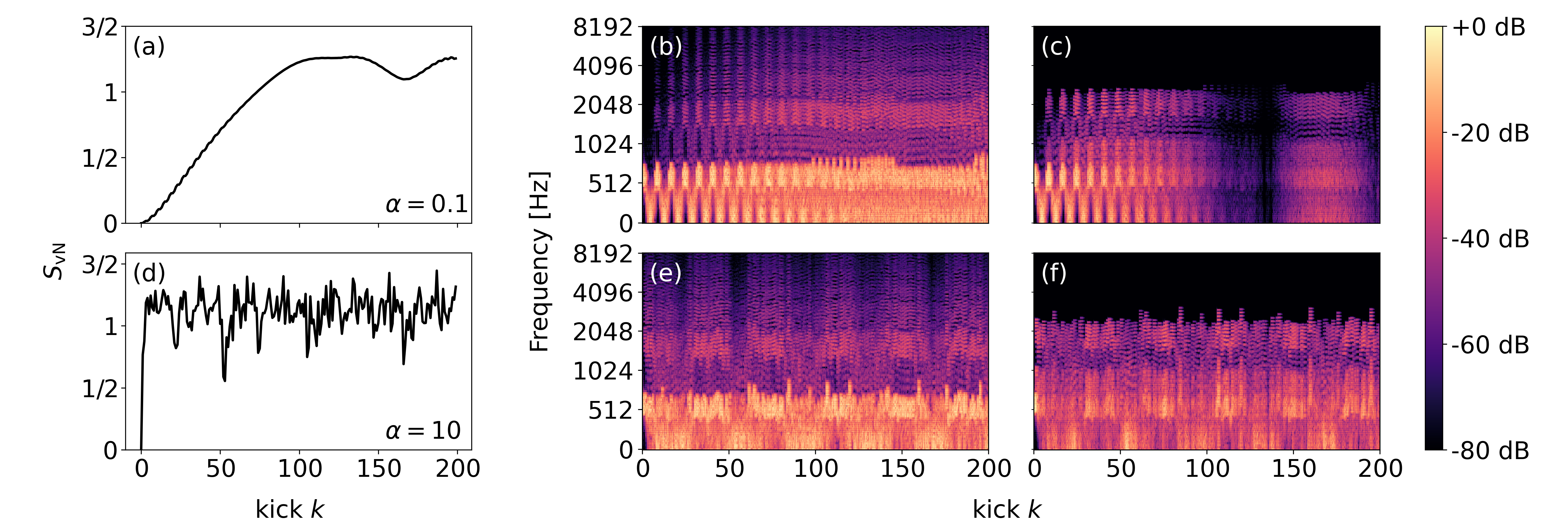}
    \caption{For the kicked rotor, at system size \(L = 8\), prepared with the initial state $|\theta, \phi\rangle=|0, 0\rangle$ for $\alpha=0.1$ [panels (a–c)] and $\alpha=10$ [panels (d–f)]. Panel labels as in Fig.~\ref{fig:svn_spectrogram_OAT}
    Animations with sound are available for
  \href{https://youtube.com/shorts/6NPSsHVl8pw}{panel~(b)},
  \href{https://youtube.com/shorts/APhLuR0DTi4}{panel~(c)},
  \href{https://youtube.com/shorts/k7_IFkN4njI}{panel~(e)},
  and \href{https://youtube.com/shorts/h3z6AAZmxQ8}{panel~(f)}.}
    \label{fig:svn_spectrogram_kicked_th0_L8}
\end{figure*}

We now turn to the Quantum Kicked Rotor, a time-periodic system whose dynamics is governed by alternating nonlinear kicks and rotations. The degree of chaotic behavior is controlled by the parameter $\alpha$, which determines the strength of the kicks. We present results for two values: $\alpha = 0.1$ (regular dynamics) and \(\alpha = 10\) (strongly chaotic regime), for the initial state \(|\theta=0, \phi=0\rangle\), and system sizes \(L=2\) and \(L=8\) (respectively Fig.~\ref{fig:svn_spectrogram_kicked_th0_L2} and Fig.\ref{fig:svn_spectrogram_kicked_th0_L8}).

For small kick strength (\(\alpha = 0.1\)), the kicked rotor exhibits regular and smooth dynamics. In the \(L = 2\) case Fig.~\ref{fig:svn_spectrogram_kicked_th0_L2}a , the entropy evolution closely resembles that of the OAT Hamiltonian, showing periodic behavior. Nevertheless, the discrete nature of the kicks is clearly visible in the spectrograms Figs.~\ref{fig:svn_spectrogram_kicked_th0_L2}b–c, and this structure is perceptibly reflected in the audio as rhythmic or pulsating modulations. The initial state modifies the evolution of both the entanglement entropy and the Husimi $Q$ function, giving rise to slightly different dynamics. Even in a nearly integrable regime, this highlights how the choice of initial state influences both the entanglement scale and the perceived sonic features. We omit the corresponding figure here, as the behavior was not qualitatively distinct from previously shown cases.
\(L = 8\) at \(\alpha = 0.1\) is characterized by gradual growth followed by slight fluctuations. A small dip in entropy is observed around kick number 170 in Fig.~\ref{fig:svn_spectrogram_kicked_th0_L8}a. This anomaly is particularly noticeable in the triangular waveform spectrogram (see Fig.~\ref{fig:svn_spectrogram_kicked_th0_L8}b), where it manifests as a brief reduction in spectral density and pitch activity. The initial state modifies how fast the growth occurs. As for \(L = 2\), we omit the corresponding figure here, as the behavior was not qualitatively distinct from previously shown cases.

In the strongly chaotic regime (\(\alpha = 10\)), the system exhibits dramatically different behavior. Regardless of initial condition or system size, the entropy becomes highly irregular and fluctuates in a seemingly random manner (see Fig.~\ref{fig:svn_spectrogram_kicked_th0_L2}d and Fig.\ref{fig:svn_spectrogram_kicked_th0_L8}d). This is reflected sonically in spectrograms (panels e–f), where the frequency components exhibit sudden shifts, producing a pitch pattern with intervallic skips—distinct from the more continuous profiles of the integrable OAT model and kicked rotor with small values of \(\alpha\).

Despite the shared chaotic structure, the system size still influences the entanglement profile. For \(L=2\), entropy remains below 1 with large fluctuations of about \(\pm 0.5\) (Fig.~\ref{fig:svn_spectrogram_kicked_th0_L2}d), whereas for \(L=8\), it rapidly rises and stabilizes near \(S_{vN} \approx 1\), fluctuating around this plateau (Fig.~\ref{fig:svn_spectrogram_kicked_th0_L8}d). This reflects faster entanglement generation and information scrambling in larger chaotic systems, where the effective Hilbert space is significantly larger.

Overall, the sonification distinguished model-specific entanglement dynamics and facilitated intuitive perception of underlying quantum behaviors. Additional sonified examples—including audio, spectrograms, von Neumann entropy plots, and videos showing quantum state evolution on the Bloch sphere and in phase space—are available for a broader range of system sizes, parameters, and Hamiltonians on the project’s GitHub repository.

\section{Discussion and Future Directions}\label{sec:discussion}

This study demonstrates the value of sonification as both a scientific and artistic tool for exploring quantum dynamics. The auditory rendering of abstract quantities such as von Neumann entropy and Husimi Q distributions provides a new channel for engaging with quantum behavior—particularly entanglement growth and the distinction between integrable and chaotic systems. By translating mathematical structures into sound, the method complements conventional visualizations and facilitates a more intuitive, temporally resolved understanding of complex quantum phenomena.

The artistic dimension of sonification not only enhances accessibility but also opens pathways to interdisciplinary collaboration. The generated audio material has potential for creative reuse in musical composition, sound design, and multimedia art. Providing access to sonified datasets through open platforms such as GitHub encourages broader engagement beyond the physics community.

Future work should also explore more immersive spatialization techniques. In particular, deploying the sonification in ambisonic or dome-based audio systems would enable spatial encoding of the Bloch sphere or Husimi Q-function data, improving the perceptual mapping between quantum phase space and auditory space. Such high-dimensional rendering could enhance listeners’ ability to internalize features like angular spread, symmetry breaking, or localization.

To rigorously evaluate the effectiveness of these methods, perceptual and cognitive studies should be conducted to assess whether sonification genuinely improves understanding of quantum behavior—especially among students or non-expert audiences. Metrics such as learning gain, pattern recognition accuracy, or intuitive classification of chaotic vs. regular systems could be used to evaluate impact.

On the technical side, one challenge we encountered was the need to downsample high-resolution quantum simulation data for audio rendering. Although this compression minimally affected perceptual clarity in our experiments, future implementations could benefit from sampling that preserve all the features. Additionally, integrating interactive controls—such as real-time parameter modulation or spatial navigation—would make the sonification experience more exploratory and user-driven, though it may come at a significant computational cost. 

Finally, extending this framework to more complex systems presents a compelling opportunity. These systems often defy easy visual interpretation, making them prime candidates for perceptually grounded approaches like sonification. In particular, the proposed sonification approach enables the analysis of complex quantum circuits on digital quantum computers by sonifying each successive circuit layer. A natural next step would be to incorporate many‑body entanglement measures—such as entanglement depth and non‑$k$‑separability based on many-body Bell correlations \cite{Plodzien2024_2}—which capture many-body entanglement structure of a given quantum state. In particular, these many-body Bell correlations characterize entanglement structure of graph states \cite{Plodzien_graph_2024}, thus combining with Husimi function analysis, would allow to provide audio-representation of a quantum states which are naturally represented as a mathematical graphs. Sonification of mixed quantum states, in particular finite temperature Gibbs state, is another exciting possibility. Using more complicating entanglement measures for sonification would reveal additional details about the state’s internal structure enhancing audio-visual content. Lastly, the proposed sonification method allows for audio-visual representation of many-body quantum Hamiltonian through sonification of its eigenstates, where the time domain is given by the eigenvalues order.

\section{Conclusion}\label{sec:conclusion}
This study establishes sonification as an intuitive method for interpreting and experiencing the dynamics of quantum entanglement. By translating abstract quantum features—such as bipartite entanglement entropy growth and Husimi function of many-qubit state—into auditory form, our approach enables both scientific insight and artistic engagement. The perceptual accessibility of sound offers a novel way to explore differences between regular and quantum chaotic dynamics, and to intuitively grasp entanglement structure in many-qubit dynamics. 

Looking ahead, the sonification of quantum entanglement—particularly in many-body and time-dependent systems—offers a promising avenue for educational and scientific communication. By making abstract quantum behaviors perceptible, this approach could support interdisciplinary efforts in physics, auditory display, and cognition. Further development of quantum-specific sonification techniques will help refine their perceptual utility and clarify their role in both research and outreach.

\section{Acknowledgments}\label{sec:acknowledgments}
\anonymize{We gratefully acknowledge the funding and support provided by ICFO for our research, as well as the Barcelona Institute of Science and Technology (BIST). 

ICFO-QOT group acknowledges support from:
European Research Council AdG NOQIA; 
MCIN/AEI (PGC2018-0910.13039/501100011033,  CEX2019-000910-S/10.13039/501100011033, Plan National FIDEUA PID2019-106901GB-I00, Plan National STAMEENA PID2022-139099NB, I00, project funded by MCIN/AEI/10.13039/501100011033 and by the “European Union NextGenerationEU/PRTR" (PRTR-C17.I1), FPI); QUANTERA DYNAMITE PCI2022-132919, QuantERA II Programme co-funded by European Union’s Horizon 2020 program under Grant Agreement No 101017733; Ministry for Digital Transformation and of Civil Service of the Spanish Government through the QUANTUM ENIA project call - Quantum Spain project, and by the European Union through the Recovery, Transformation and Resilience Plan - NextGenerationEU within the framework of the Digital Spain 2026 Agenda;
Fundació Cellex;
Fundació Mir-Puig; 
Generalitat de Catalunya (European Social Fund FEDER and CERCA program;
Barcelona Supercomputing Center MareNostrum (FI-2023-3-0024); 
Funded by the European Union. Views and opinions expressed are however those of the author(s) only and do not necessarily reflect those of the European Union, European Commission, European Climate, Infrastructure and Environment Executive Agency (CINEA), or any other granting authority.  Neither the European Union nor any granting authority can be held responsible for them (HORIZON-CL4-2022-QUANTUM-02-SGA  PASQuanS2.1, 101113690, EU Horizon 2020 FET-OPEN OPTOlogic, Grant No 899794, QU-ATTO, 101168628),  EU Horizon Europe Program (This project has received funding from the European Union’s Horizon Europe research and innovation program under grant agreement No 101080086 NeQSTGrant Agreement 101080086 — NeQST); 
ICFO Internal “QuantumGaudi” project}

\bibliographystyle{unsrt}
\bibliography{biblio_2025}

\end{document}